\documentclass[prd,aps,nofootinbib,showpacs,showkeys]{revtex4}

\def\XXint#1#2#3{{\setbox0=\hbox{$#1{#2#3}{\int}$}
     \vcenter{\hbox{$#2#3$}}\kern-.5\wd0}}

\usepackage{graphicx}
\usepackage{bm}
\usepackage{amssymb}
\usepackage{amsmath}
\usepackage{euscript}

\begin{document}

\title{Quantum fluctuations of Coulomb Potential as a Source of Flicker
Noise. The Influence of Heat Bath}

\author{Kirill~A.~Kazakov}\email{kirill@phys.msu.ru}

\affiliation{Department of Theoretical Physics,
Physics Faculty,\\
Moscow State University, $119899$, Moscow, Russian Federation}

\begin{abstract}
The power spectrum of finite-temperature quantum electromagnetic
fluctuations produced by elementary charge carriers under the
influence of external electric field is investigated. It is found
that under the combined action of the photon heat bath and the
external field, the low-frequency asymptotic of the power spectrum
is modified both qualitatively and quantitatively. The new term in
the power spectrum is inversely proportional to frequency, but is
odd with respect to it. It comes from the connected part of the
correlation function, and is related to the temperature and
external field corrections to the photon and charge carrier
propagators. In application to the case of a biased conducting
sample, this term gives rise to a contribution to the voltage
power spectrum which is proportional to the absolute system
temperature, the charge carrier mobility, the bias voltage
squared, and a factor describing dependence of the noise intensity
on the sample geometry. It is verified that the derived expression
is in agreement with the experimental data on $1/f$-noise
measurements in metal films. It is shown also that the obtained
result provides a natural resolution to the problem of divergence
of the total noise power.
\end{abstract}

\maketitle

\section{Introduction}

The origin of flicker noise \cite{buck} observed in virtually all
conducting media remains an open issue in condensed matter theory.
Despite numerous models suggested since its discovery eighty years
ago there is presently no consistent theory which would explain
the main characteristic properties of this omnipresent noise. The
power spectral density of flicker noise is proportional to
$f^{-\gamma},$ where $f$ is frequency, and the exponent $\gamma$
is about unity (usually $\gamma$ takes values $0.9$ to $1.5$). One
of the essential difficulties for theoretical explanation is the
fact that experiments show no limits for this dispersion law,
neither lower or upper. Although flicker noise dominates only at
sufficiently low frequencies, the $1/f$-component is detected in
the whole measured band up to $10^6 {\rm Hz}.$ It is also
experimentally established that the noise power spectrum of a
sample is proportional to the applied bias squared, and roughly
inversely proportional to its volume.

Ubiquity of flicker noise and universality of its properties
suggest existence of a simple reason for its occurrence. It is
natural to expect this reason to have a quantum origin. Although
some of the models suggested so far do consider various quantum
effects as underlying mechanisms of flicker noise (such as, for
instance, trapping of charge carriers), it may well be that its
origin is to be sought at the most fundamental level. Namely, it
is plausible that the phenomenon of flicker noise has its roots in
the very quantum nature of interaction of elementary charges with
electromagnetic field. From this point of view, the problem was
attacked by Handel \cite{handel1}, who suggested that the observed
flicker noise is related to the spectrum of low-energy photons
emitted in any scattering process, which, according to Handel, has
the $1/f$ profile and reflects the well-know property of
bremsstrahlung, namely, the infrared divergence of the
cross-section considered as a function of the energy loss. Later,
the argument was modified and the so-called coherent quantum $1/f$
effect described \cite{handel2} on the basis of quantum
electrodynamic results of Kibble and Zwanziger \cite{zwan,kibble}.
Although Handel's theory was severely criticized in many respects
\cite{tremblay,kampen}, it has found support in independent
investigations of Refs.~\cite{vliet,ziel}. Handel's approach is
based on consideration of {\it current} fluctuations. An
essentially different quantum approach to the problem was recently
proposed \cite{kazakov1,kazakov2}, in which flicker noise is
treated as a {\it voltage} fluctuation originating from quantum
fluctuations of individual electric fields of charge carriers. In
this case, the noise power spectrum is found by evaluating the
two-point correlation function of the Coulomb potential of an
elementary particle, dispersion of this function being related to
the quantum spreading of the particle wave packet. However, the
magnitude of noise induced by an external electric field, given by
this theory, turned out to be too small to explain the observed
noise level.

The purpose of this paper is to show that the value of induced
quantum noise is actually significantly higher than the previously
calculated. It turns out that in evaluating the effect of external
electric field it is essential to take into account statistical
properties of electromagnetic field. The point is that
simultaneous account of the effects of photon heat bath and
external field leads to appearance of a contribution of new type.
Namely, the new term in the noise power spectrum is {\it odd} with
respect to frequency, and hence corresponds to the part of the
correlation function, which is odd with respect to the difference
of its time arguments. The underlying reason that makes the
appearance of this term possible is the inhomogeneity in time of
fluctuations produced by individual charge carriers. As a
consequence of this inhomogeneity, the correlation function of
each individual contribution to the Coulomb field fluctuation
depends on both time arguments separately, rather than on their
difference only, so that symmetry under interchanging of the
arguments does not forbid appearance of the odd contribution. The
time homogeneity is restored only after summing up all independent
contributions. The new contribution stems from the connected part
of the correlation function of particle's field, rather than from
the disconnected one that was in focus of Ref.~\cite{kazakov2},
and is related to the temperature and external field corrections
to the photon and charge carrier propagators, in contrast to
considerations of Ref.~\cite{kazakov2} where thermal bath and
external field affected only real particle states.

The paper is organized as follows. In Sec.~\ref{prelim1} we
briefly discuss the role of the heat bath effects in evaluating
the mean value and the correlation function of electromagnetic
field produced by elementary particles, and identify the
contributions relevant in the low-frequency regime. The power
spectral functions of the Coulomb field and voltage fluctuations
are defined in Sec.~\ref{prelim2}, and written in the form
convenient for explicit calculations which are carried out in
Sec.~\ref{calcul}. The low-frequency asymptotic of the power
spectrum of Coulomb potential fluctuations in the absence of
external electric field is evaluated in Sec.~\ref{zerofield}, and
is found to exhibit an inverse frequency dependence. However, the
$1/f$ part of the spectrum is cancelled in the expression for the
voltage spectral density. The non-vanishing $1/f$-contribution to
the voltage power spectrum is obtained in Sec.~\ref{nonzerofield}
upon account of the influence of external homogeneous electric
field on the virtual charger particle propagation. Application of
the obtained results to solids and comparison with experimental
data is given in Sec.~\ref{comparison}. Gauge independence of our
treatment of electromagnetic fluctuations is proved in the
Appendix.

\section{Preliminaries}

\subsection{Heat bath contribution to particle propagators}\label{prelim1}

Consider a quantized system of charged particles interacting with
electromagnetic field. Let $T$ be the absolute temperature of the
system. We are interested in the influence of finite temperature
on quantum properties of the electromagnetic field produced by
charged particles. Specifically, finite-temperature correlations
in the values of the particle's Coulomb fields will be
investigated. To the leading order in the electromagnetic
coupling, these correlations are a single-particle effect, in the
sense that in this case only fields produced by one and the same
particle correlate. In what follows, we thus confine ourselves to
systems which allow perturbative treatment of charge carrier
collisions, either in terms of original particles, or in terms of
quasi-particles (e.g., conduction electrons in metals). In the
latter case, the particle's mass and energy-momentum relation
should be replaced by the effective ones.

In this section, we shall discuss some general features of the
temperature effect, related to the heat bath influence on virtual
propagation of the electromagnetic and charged field quanta.
Evidently, the heat bath has no effect on the mean electromagnetic
field of a charged particle. Indeed, under above assumptions, this
field can be represented as the amplitude of one-photon emission
in a transition between free charged particle states, contracted
with the photon propagator. Since the 4-vector of momentum
transfer to a free massive particle, $p,$ is always spacelike,
$p^2<0,$ distribution of real photons appearing in the definition
of the photon propagator is immaterial in the calculation of the
field. As a consequence, quantities built from the mean field,
such as the disconnected part of the correlation function, are not
affected by the photon heat bath (details of evaluation of this
part see in Refs.~\cite{kazakov1,kazakov2}). Things change,
however, when the connected part of the two-point correlation
function of electric potential is considered. It is defined by the
following symmetric expression
\begin{eqnarray}\label{corr}&&
C^{\rm con}_{00}(x;x') = \frac{1}{2}\langle {\rm
in}|\hat{A}_0(x)\hat{A}_0(x') + \hat{A}_0(x')\hat{A}_0(x) |{\rm
in}\rangle\,,
\end{eqnarray}
\noindent where $x$ and $x'$ are the spacetime coordinates of two
observation points, $\hat{A}_0$ is the scalar potential Heisenberg
operator, and $|{\rm in }\rangle$ denotes the given {\it in} state
of the system ``charged particle + electromagnetic field.'' In the
two-photon processes, photon momenta are allowed to take on
lightlike directions, and hence the photon heat bath does
contribute to the function $C^{\rm con}_{00}(x;x').$

As is well-known, the ordinary Feynman rules of the S-matrix
theory are not generally applicable for the calculation of {\it
in-in} expectation values, and must be modified, e.g., according
to Schwinger and Keldysh \cite{keldysh}. This complication was
overcome in \cite{kazakov2} by rewriting Eq.~(\ref{corr}) in the
form
\begin{eqnarray}\label{corr2}&&
C^{\rm con}_{00}(x;x') = {\rm Re}\langle {\rm out}|T
\{\hat{A}_0(x)\hat{A}_0(x')\}|{\rm in}\rangle\,,
\end{eqnarray}
\noindent which allows the use of the S-matrix rules. This
transformation uses equivalence of the one-particle {\it in} and
{\it out} states (and Hermiticity of the electromagnetic field
operator). It is applicable to the present case as well, despite
the fact that now the {\it in} state is not one-particle, because
we are not going to take into account scattering processes in the
heat bath itself. This can also be shown directly at the
diagrammatic level following the route of transformations taken in
Ref.~\cite{kazakov}, which is formally the same in zero- and
nonzero-temperature cases. Thus, in order to calculate the {\it
in-in} expectation value (\ref{corr}) taking into account the heat
bath effect, the standard finite-temperature-field-theory
techniques can be used \cite{landsman}. Below we employ a version
of the real time formalism, developed in \cite{niemi}, which is
especially convenient in actual calculations since the momentum
space propagators in this formulation do not involve the step
function.

The real time formulation involves doubling of all fields, which
will be specified by a two-valued lower index. According to the
diagrammatic rules derived in \cite{niemi}, the photon propagator
has the following matrix structure\footnote{The photon propagator
is written here in the Feynman gauge. The question of gauge
independence of the correlation function is considered in the
Appendix.}
\begin{eqnarray}\label{phiphot}
\mathfrak{D}_{\mu\nu}(x) &=& \int\frac{d^4 k}{(2\pi)^4}
\mathfrak{D}_{\mu\nu}(k)e^{-ikx}\,, \quad \mathfrak{D}_{\mu\nu}(k)
= 4\pi\eta_{\mu\nu}\left(
\begin{array}{cc}
D_{11}(k)&D_{12}(k)\\
D_{21}(k)&D_{22}(k)
\end{array}\right)\,,
\end{eqnarray}
\noindent where $$D_{11}(k) = - D^*_{22}(k) = \frac{1}{k^2 + i0} -
\frac{2\pi i \delta(k^2)}{e^{\beta|k_0|} - 1}\,, \quad D_{12}(k) =
D_{21}(k) = - \frac{2\pi
i\delta(k^2)e^{\beta|k_0|/2}}{e^{\beta|k_0|} - 1}\,,$$ $\beta =
1/T$ being the inverse absolute temperature of the system. In
applications to the problem of $1/f$-noise considered below, the
value of the product $\beta |k_0|$ turns out to be very small. For
instance, even for frequencies as large as $10^6 {\rm Hz},$ and
temperatures as small as $1^{\circ}{\rm K}, $ it is less than
$10^{-27}\cdot 10^6/(10^{-16}) = 10^{-5}$ (the factors $10^{-27}$
and $10^{-16}$ are contributed by the Planck and Boltzmann
constants, respectively), so the denominators in the above
expressions can be replaced by $\beta |k_0|,$ implying that the
second term in $D_{11}$ dominates. On the other hand, the
temperature effect on the propagation of massive particles is much
less prominent. For instance, in the case of conduction electrons
in a crystal (this case will be used throughout as a standard
example), the particle energy $\varepsilon$ is of the order
$(\hbar/d)^2/m,$ where $m$ is the particle mass, and $d$ is the
lattice spacing. Taking $d\sim 10^{-8}{\rm cm},$ $m\sim
10^{-27}{\rm g},$ we find that $\varepsilon/T \sim 10^{+5},$ so
that the temperature contributions can be completely neglected
(for fermions as well as for bosons), and the propagator taken in
the simple diagonal form
\begin{eqnarray}\label{phiprop}
\mathfrak{D}^{\phi}(x) &=& \int\frac{d^4 k}{(2\pi)^4}
\mathfrak{D}^{\phi}(k)e^{-ikx}\,, \quad \mathfrak{D}^{\phi}(k) =
\left(
\begin{array}{cc}
D^{\phi}_{11}(k)& 0\\
0& D^{\phi}_{22}(k)
\end{array}\right)\,, \\
D^{\phi}_{11}(k) &=& - D^{\phi *}_{22}(k) = (m^2 - k^2 -
i0)^{-1}\,.\nonumber
\end{eqnarray}
\noindent This is for a scalar particle described by the action
\begin{eqnarray}\label{action}
S_{\phi} &=& {\displaystyle\int} d^4 x
\left\{(\partial_{\mu}\phi^* + i e A_{\mu}\phi^*)
(\partial^{\mu}\phi - ie A^{\mu}\phi) - m^2 \phi^*\phi\right\}\,,
\end{eqnarray} \noindent where $e$ is the particle charge.
Account of particle spin, though adds some extra algebra, does not
change the long-range properties of its field, so that following
Ref.~\cite{kazakov2} we work with the simplest case of zero-spin
particles. The matrix propagators are multiplied in the
interaction vertices, generated by the triple and higher order
terms in the Lagrangian, with an additional minus sign for the
product of 2-components, as in the Schwinger-Keldysh techniques.
The arguments of the Green functions are treated as 1-component
fields. Finally, external particle lines represent normalized
particle amplitudes or their conjugates, according to whether the
particle is incoming or outgoing, just like in the conventional
techniques.

As was mentioned above, the photon propagator is dominated by the
temperature contribution (as long as the range of momentum
integration contains lightlike directions, see discussion in
Sec.~\ref{zerofield}), while in the massive particle propagator
this contribution is negligible. It is important, on the other
hand, that the heat bath affects significantly the real particle
propagation, i.e., external matter lines in the diagrams.
Bilinears of the particle amplitudes representing these lines are
expressed eventually via statistical distribution function (see
Sec.~\ref{nonzerofield} for details). Thus, apart from explicit
$T$-dependence coming from the photon propagator, the correlation
function also depends on temperature implicitly through the
particle statistical distribution.

\subsection{Power spectral densities of potential and voltage fluctuations}
\label{prelim2}

Connected contribution to the power spectral density of electric
potential fluctuations is obtained by Fourier transforming
Eq.~(\ref{corr}) with respect to the difference of the time
instants $t,t'$:
\begin{eqnarray}\label{corrf}&&
C(\bm{x},\bm{x}',t',\omega) = \int\limits_{-\infty}^{+\infty}d\tau
C^{\rm con}_{00}(\bm{x},t'+\tau;\bm{x}',t')e^{-i\omega\tau}\,.
\end{eqnarray}
\noindent The upper and lower indices in the notation of the
correlation function are suppressed in the left hand side, for
brevity. We are interested ultimately in the power spectrum of
voltage fluctuations, $C_U,$ measured between two observation
points $\bm{x},\bm{x}'$ (e.g., two leads attached to a conducting
sample). The connected contribution to the voltage correlation
function is given by
\begin{eqnarray}\label{corru}&&
C_U(\bm{x},\bm{x}',t,t') = \frac{1}{2}\langle{\rm
in}|\hat{U}(t)\hat{U}(t') + \hat{U}(t')\hat{U}(t)|{\rm
in}\rangle\,,
\end{eqnarray}
\noindent where $\hat{U}(t) = \hat{A}_0(\bm{x},t) -
\hat{A}_0(\bm{x}',t)$ is the operator of voltage between the two
points. This function is separately symmetric with respect to the
interchanges $\bm{x}\leftrightarrow \bm{x}',$ and
$t\leftrightarrow t',$ unlike the function $C^{\rm
con}_{00}(x;x')$ which is only symmetric under $x\leftrightarrow
x'.$ Substituting the definition of $\hat{U}(t)$ in
Eq.~(\ref{corru}), the former can be expressed via the latter
\begin{eqnarray}
C_U(\bm{x},\bm{x}',t,t') &=& C^{\rm con}_{00}(\bm{x},t;\bm{x},t')
+ C^{\rm con}_{00}(\bm{x}',t;\bm{x}',t') \nonumber \\ &-&
\left[C^{\rm con}_{00}(\bm{x},t;\bm{x}',t') + C^{\rm
con}_{00}(\bm{x}',t;\bm{x},t')\right]\,.
\end{eqnarray}
\noindent Accordingly, the power spectral density of voltage
fluctuations, defined by
\begin{eqnarray}\label{corruf}&&
C_U(\bm{x},\bm{x}',t',\omega) =
\int\limits_{-\infty}^{+\infty}d\tau
C_U(\bm{x},t'+\tau,\bm{x}',t')e^{-i\omega\tau}\,,
\end{eqnarray}
\noindent is expressed through that of potential fluctuations as
\begin{eqnarray}\label{corrfup}
C_U(\bm{x},\bm{x}',t',\omega) &=& C(\bm{x},\bm{x},t',\omega) +
C(\bm{x}',\bm{x}',t',\omega) - \left[C(\bm{x},\bm{x}',t',\omega) +
C(\bm{x}',\bm{x},t',\omega)\right]\,.
\end{eqnarray}
\noindent Although $C_U(\bm{x},\bm{x}',t,t')$ is symmetric with
respect to the interchange $t\leftrightarrow t',$ it depends on
both time arguments separately, and therefore
$C_U(\bm{x},\bm{x}',t,\omega)$ does not have to be an even
function of $\omega.$

When calculating the power spectrum of potential fluctuations
according to Eqs.~(\ref{corr2}), (\ref{corrf}), it is convenient
to perform the Fourier transformation under the sign ``{\rm Re}''
in Eq.~(\ref{corr2}). For this purpose we introduce the Fourier
transform of the two-point Green function:
\begin{eqnarray}\label{fourierg}&&
G(\bm{x},\bm{x}',t',\omega) = \int\limits_{-\infty}^{+\infty}d\tau
G(\bm{x},t'+\tau;\bm{x}',t') e^{-i\omega\tau}\,, \quad G(x;x') =
\langle {\rm out}|T \{\hat{A}_0(x)\hat{A}_0(x')\}|{\rm
in}\rangle\,, \nonumber \\
\end{eqnarray}
\noindent with the help of which the power spectral density of
potential fluctuations can be written as
\begin{eqnarray}\label{corrg}
C(\bm{x},\bm{x}',t',\omega) &=& \frac{1}{2}{\rm Re
}\left\{G(\bm{x},\bm{x}',t',\omega) +
G(\bm{x},\bm{x}',t',-\omega)\right\} \nonumber\\ &+&
\frac{i}{2}{\rm Im }\left\{G(\bm{x},\bm{x}',t',\omega) -
G(\bm{x},\bm{x}',t',-\omega)\right\}\,.
\end{eqnarray}
\noindent We see that contributions to the function
$C(\bm{x},\bm{x}',t',\omega),$ and hence to the voltage power
spectrum, are either real even, or imaginary odd functions of
frequency.

\section{Evaluation of low-frequency asymptotic of spectral density}\label{calcul}

\subsection{Power spectrum in the absence of external electric field}\label{zerofield}

\begin{figure}
\includegraphics{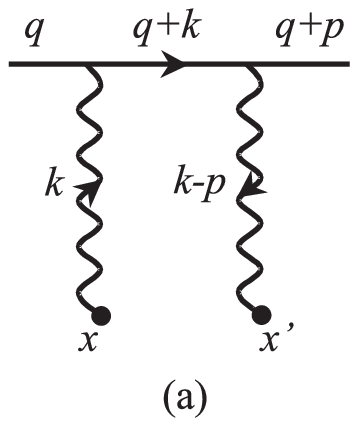}
\includegraphics{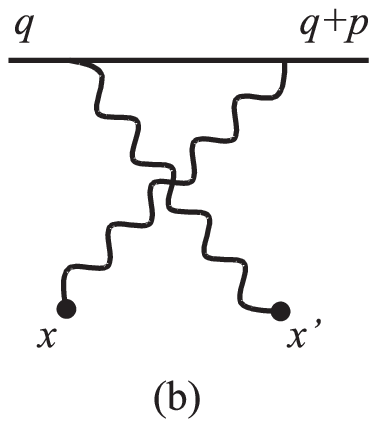}
\includegraphics{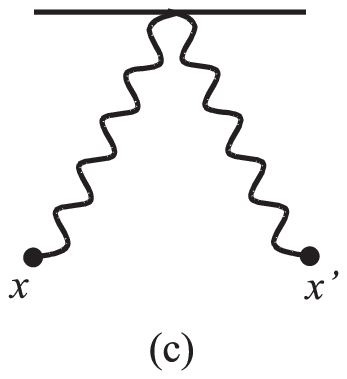}
\caption{Kazakov, Quantum fluctuations of Coulomb potential as a
source of flicker noise. The influence of heat bath} \label{fig1}
\end{figure}

The tree contribution to the right hand side of Eq.~(\ref{corr2})
when the influence of external electric field is neglected is
shown in Fig.~\ref{fig1}. Repeating the argument of
Ref.~\cite{kazakov2}, it is not difficult to show that the leading
contribution is contained in the diagrams \ref{fig1}(a),
\ref{fig1}(b). Distinguishing their contributions by the
corresponding Latin subscript, we have
\begin{eqnarray}\label{diagen}
G_a(x,x') &=& ie^2\iint d^4 z d^4 z'
\left\{\mathfrak{D}(x,z)\left[\phi_0(z)
\stackrel{\leftrightarrow}{\partial_{0}}
\mathfrak{D}^{\phi}(z,z')\stackrel{\leftrightarrow}{\partial_{0}^{\,\prime}}
\phi^*_0(z')\right]\mathfrak{D}(z',x')\right\}_{11}\,, \\
G_b(x,x') &=& G_a(x',x)\,,\nonumber
\end{eqnarray}
\noindent where
\begin{eqnarray}
\varphi\stackrel{\leftrightarrow}{\partial_{0}}\psi &=&
\varphi\partial_{0}\psi - \psi\partial_{0}\varphi\,, \nonumber
\end{eqnarray} and $\phi_0$ is the given particle state.
\noindent Going over to momentum space with the help of
Eqs.~(\ref{phiphot}), (\ref{phiprop}), introducing the spectral
function for $G_a(x,x')$ according to Eq.~(\ref{fourierg}), and
writing the matrix product longhand yields
\begin{eqnarray}\label{diagenk1}
G_a(\bm{x},\bm{x}',t',\omega) &=& (4\pi e)^2\iint \frac{d^3
\bm{q}}{(2\pi)^3} \frac{d^3
\bm{p}}{(2\pi)^3}\frac{a(\bm{q})a^*(\bm{q} +
\bm{p})}{\sqrt{2\varepsilon_{\bm q}2\varepsilon_{{\bm q} +
\bm{p}}}} e^{ip^0(t' - t_0) - i\bm{p}\bm{x}'}
J_a(p,q,\bm{x}-\bm{x}',\omega)\,, \nonumber \\
p^0 &=& \varepsilon_{\bm{q} + \bm{p}} - \varepsilon_{\bm{q}}\,,
\quad \varepsilon_{\bm{q}} = +\sqrt{\bm{q}^2 + m^2}\,,
\end{eqnarray}
\noindent where
\begin{eqnarray}\label{diagenk3}&&
J_a(p,q,\bm{x}-\bm{x}',\omega) =  - i\int \frac{d^3
\bm{k}}{(2\pi)^3} e^{i\bm{k}(\bm{x}'-\bm{x})}(2q_0 + k_0)(2q_0 +
k_0 + p_0)\nonumber\\&&\times
\left[D_{11}(k)D^{\phi}_{11}(q+k)D_{11}(k-p) +
D_{12}(k)D^{\phi}_{22}(q+k)D_{21}(k-p)  \right]_{k_0=\omega}\,.
\end{eqnarray}
\noindent Here $q_{\mu}$ is the particle 4-momentum, $a(\bm{q})$
its momentum wave function at some time instant $t_0,$ normalized
by
\begin{eqnarray}\label{norm}
\int\frac{d^3 \bm{q}}{(2\pi)^3}|a(\bm{q})|^2 = 1\,,
\end{eqnarray}\noindent and it is taken into account that $D^{\phi}_{12} = 0\,.$

The second term in the square brackets in Eq.~(\ref{diagenk3}) can
be neglected. Indeed, in view of the factor $D_{12}(k)$ which is
proportional to $\delta(k^2),$ and the condition $k_0 = \omega,$
the momentum $k$ contributes only a tiny value to the argument of
the factor $D_{21};$ for electrons in a crystal, for instance, the
ratio $|\bm{k}|/|\bm{p}|$ is of the order $(\hbar
\omega/c)/(\hbar/d) = \omega d/c \approx \omega \cdot
10^{-18}\,{\rm s}.$ Therefore, this factor can be written simply
as $D_{21}(p)\sim \delta(p^2).$ On the other hand, momentum
transfer $p$ to the massive particle is spacelike, and hence the
argument of the delta-function is always nonzero. Furthermore,
using explicit expression for the photon propagator, the first
term in the square brackets reads
\begin{eqnarray}&&
\hspace{-0,5cm}\left\{\frac{1}{k^2 + i0} - \frac{2\pi i
\delta(k^2)}{e^{\beta|k_0|} - 1} \right\}D^{\phi}_{11}
\left\{\frac{1}{(k-p)^2 + i0} - \frac{2\pi i
\delta((k-p)^2)}{e^{\beta|k_0-p_0|} - 1} \right\} = \frac{1}{[k^2
+ i0]}D^{\phi}_{11}\frac{1}{[(k-p)^2 + i0]} \nonumber\\&& -
\frac{2\pi i \delta(k^2)}{e^{\beta|k_0|} - 1}D^{\phi}_{11}
\frac{1}{(k-p)^2 + i0} - \frac{1}{k^2 + i0}D^{\phi}_{11}\frac{2\pi
i \delta((k-p)^2)}{e^{\beta|k_0-p_0|} - 1} + \frac{2\pi i
\delta(k^2)}{e^{\beta|k_0|} - 1}D^{\phi}_{11}\frac{2\pi i
\delta((k-p)^2)}{e^{\beta|k_0-p_0|} - 1}\,.\nonumber
\end{eqnarray}
\noindent As before, the last term on the right hand side can be
neglected, while the first term, describing the zero-temperature
contribution, has already been considered in Ref.~\cite{kazakov2}.
Furthermore, $k_0$ enters the temperature exponent in the third
term in the combination $(k_0 - p_0),$ and, therefore, this term
does not contribute to the leading order in $1/\omega,$ because in
practice $|\omega|\ll p_0.$ Indeed, estimating the energy transfer
as $p_0 \approx (\bm{p}\bm{q})/m,$ and taking $|\bm{p}| \sim 1/l,$
where $l$ is the characteristic sample length, one finds for our
standard example $|\omega/p_0|\sim 10^{-8}|\omega|l,$ where
$l,\omega$ are supposed to be expressed in the $CGS$ system of
units. Even for $l$ as large as $1\,{\rm cm},$ this ratio is very
small for all practically relevant frequencies. Thus, the
contribution of the second term only remains to be considered. In
view of the factor $\delta(k^2),$ the pole of the function
$D_{11}(q+k)$ in this term does not contribute. This is again a
consequence of the requirement that the momentum transfer to a
massive particle on-shell be spacelike: conditions $k^2 = 0,$ $q^2
= m^2,$ and $(q+k)^2 = m^2$ cannot be satisfied altogether. Hence,
the scalar particle propagator can be written simply as
$D_{11}(q+k) = - 1/(2qk),$ or, assuming that the particle is
nonrelativistic, $|\bm{q}|\ll m,$ as
\begin{eqnarray}\label{prsingular}
D^{\phi}_{11}(q+k) = -\frac{1}{2m\omega}\,.
\end{eqnarray}
\noindent Finally, neglecting $k$ in comparison with $p$ in the
factor $1/[(k-p)^2 + i0],$ using $p^2<0$ to omit $i0,$ and
retaining only terms singular in $\omega,$ we find
\begin{eqnarray}\label{diagenk4}
J_a(p,q,\bm{x}-\bm{x}',\omega) &=& \frac{2\pi (2m +
p_0)}{\omega\beta |\omega| p^2}\int \frac{d^3 \bm{k}}{(2\pi)^3}
e^{i\bm{k}(\bm{x}'-\bm{x})}\delta(\omega^2 - \bm{k}^2 )
\nonumber\\
&=& \frac{(2m + p_0)}{2\pi \omega\beta p^2} + O(\omega)\,.
\end{eqnarray}
\noindent The reason why $p_0$ has been kept along with $m$ in the
numerator will become clear soon. As to the contribution of the
diagram \ref{fig1}(b), changing $\bm{k} \to \bm{k} + \bm{p}$ in
Eq.~(\ref{diagenk3}), and then $\bm{q}\to \bm{q} - \bm{p},$
$\bm{p}\to -\bm{p}$ in Eq.~(\ref{diagenk1}) shows that
$$G_b(\bm{x},\bm{x}',t',\omega) = G^*_a(\bm{x},\bm{x}',t',-\omega)\,.$$
Thus, the total contribution to the function
$G(\bm{x},\bm{x}',t',\omega)$ is
\begin{eqnarray}\label{gab}
G(\bm{x},\bm{x}',t',\omega) &=& G_a(\bm{x},\bm{x}',t',\omega) +
G^*_a(\bm{x},\bm{x}',t',-\omega)\,.
\end{eqnarray}
\noindent It is seen from this relation and Eq.~(\ref{diagenk4})
that only imaginary part of $G_a$ gives rise to a nonzero
contribution to the total Green function $G,$ and this part
corresponds to the term proportional to $p_0$ on the right hand
side of Eq.~(\ref{diagenk4}). Thus,
\begin{eqnarray}\label{diagenk5}
G(\bm{x},\bm{x}',t',\omega) &=& - \frac{8\pi e^2}{m^2\beta
\omega}\iint \frac{d^3 \bm{q}}{(2\pi)^3} \frac{d^3
\bm{p}}{(2\pi)^3}a(\bm{q})a^*(\bm{q}+\bm{p})\frac{(\bm{p}\bm{q})}{\bm{p}^2}e^{
ip_0(t' - t_0) - i\bm{p}\bm{x}'}\,,
\end{eqnarray}
\noindent where, in the denominator, the energies
$\varepsilon_{\bm{q}},\varepsilon_{\bm{q}+\bm{p}}$ have been
replaced by $m,$ and $p^0$ neglected in comparison with $|\bm{p}|$
on account of the condition $|\bm{q}|\ll m,$ while in the
numerator, $p^0$ has been replaced by its leading long-range term,
$(\bm{p}\bm{q})/m.$ It is instructive to see what the right hand
side of Eq.~(\ref{diagenk5}) becomes in a particulary simple model
case when the amplitude $a(\bm{q})$ can be written as
\begin{eqnarray}\label{abrel}
a(\bm{q}) = b(\bm{q})e^{-i\bm{q}\bm{x}_0}\,,
\end{eqnarray}\noindent
where $b(\bm{q})$ is a real function of the particle momentum. In
this case, $\bm{x}_0$ is easily identified as the mean particle
position, and hence, the function $b(\bm{q})$ describes the
momentum space profile of the particle wave packet. After
extraction of the position-dependent phase factor, the amplitude
becomes a relatively slowly varying function of the particle
momentum, therefore, to the leading order of the long-range
expansion, $b(\bm{q} + \bm{p})$ in the integrand of
Eq.~(\ref{diagenk5}) can be replaced by $b(\bm{q})$
\begin{eqnarray}\label{diagenk51}
G(\bm{x},\bm{x}',t',\omega) &=& - \frac{8\pi e^2}{m^2\beta
\omega}\iint \frac{d^3 \bm{q}}{(2\pi)^3} \frac{d^3
\bm{p}}{(2\pi)^3}|b(\bm{q})|^2\frac{(\bm{p}\bm{q})}{\bm{p}^2}e^{
ip_0(t' - t_0) + i\bm{p}(\bm{x}'-\bm{x}_0)}\,,
\end{eqnarray}
\noindent In the exponent, $p_0$ generally cannot be replaced by
$(\bm{p}\bm{q})/m,$ because the subleading term of its long-range
expansion, $\bm{p}^2/(2m),$ although small compared to
$(\bm{p}\bm{q})/m,$ can change the phase significantly, provided
that the difference $(t'-t_0)$ is sufficiently large, as is the
case if the particle collisions are neglected. However, taking
into account the latter reduces this difference to the particle
mean free time, $\tau_f,$ so that the product $p_0(t'-t_0)$ can be
completely neglected. Indeed, $\tau_f$ can be estimated roughly as
$d/(|\bm{q}|/m),$ and hence, $p_0(t'-t_0)\sim
|\bm{p}||\bm{q}|\tau_f/m \sim d|\bm{p}| \sim
d/|\bm{x}'-\bm{x}_0|\ll 1.$ Then the $\bm{p}$ integral can be
evaluated with the help of the formula
$$\int\frac{d^3\bm{p}}{(2\pi)^3}e^{i(\bm{p}\bm{x})}\frac{4\pi\bm{p}}{\bm{p}^2}
= \frac{i \bm{x}}{|\bm{x}|^3}\,,$$ yielding
\begin{eqnarray}\label{gzero}
G(\bm{x},\bm{x}',t',\omega) &=& - 2i\frac{e^2}{m^2\beta
\omega}\frac{(\overline{\bm{q}}\bm{r}')}{r'^3}\,, \quad \bm{r}' =
\bm{x}' - \bm{x}_0\,,
\end{eqnarray}
\noindent where the overline denotes $\bm{q}$-averaging over the
given particle state. In the absence of external electric field,
$\overline{\bm{q}}$ is zero, and therefore so is the right hand
side of Eq.~(\ref{gzero}). As we will see in the next section, the
same result is obtained in the general case without the use of the
model decomposition (\ref{abrel}). But even for nonzero
$\overline{\bm{q}},$ the voltage power spectrum calculated from
$G$ given by Eq.~(\ref{gzero}) turns out to be zero. This is
verified directly by substituting expression (\ref{gzero}) into
Eqs.~(\ref{corrg}), (\ref{corrfup}). The reason for nullification
of the voltage power spectrum is easily identified -- it is the
consequence of the fact that the function
$G_a(\bm{x},\bm{x}',t',\omega)$ is independent of the $\bm{x}$
coordinate. The $\bm{x}$-dependence has been lost upon extracting
the low-frequency asymptotic of $J_a$ in Eq.~(\ref{diagenk4}).

\subsection{The influence of external electric field and particle
collisions}\label{nonzerofield}

\begin{figure}
\includegraphics{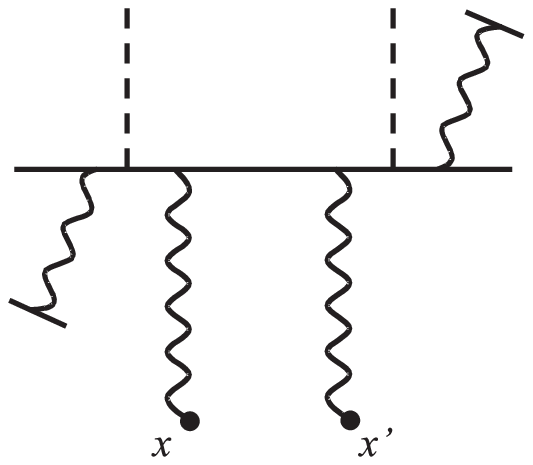}
\caption{Kazakov, Quantum fluctuations of Coulomb potential as a
source of flicker noise. The influence of heat bath} \label{fig2}
\end{figure}

Let us now consider corrections to the power spectrum due to
constant homogeneous external electric field, taking into account
also the influence of particle collisions. These corrections are
twofold. First of all, the field affects the particle wave
function $a(\bm{q}),$ which is symbolized in Fig.~\ref{fig2} by
inserting the vertices of particle-field interaction into the two
external solid lines. This is only a schematic picture, because
the effect of a constant homogeneous field on the free particle
states cannot be treated perturbatively. The latter circumstance,
however, is not important in view of the particle collisions which
prevent the particle from gaining too much momentum from the
field, thus cutting down its effect (particle collisions are
symbolized in Fig.~\ref{fig2} by a virtual photon interchange
between particles). Account of these two factors is accomplished
by replacing the particle momentum probability distribution,
$|a(\bm{q})|^2,$ by the statistical distribution function,
obtained as a solution of the kinetic equation in the presence of
external electric field. This point will be discussed in more
detail later in this section.

\begin{figure}
\includegraphics{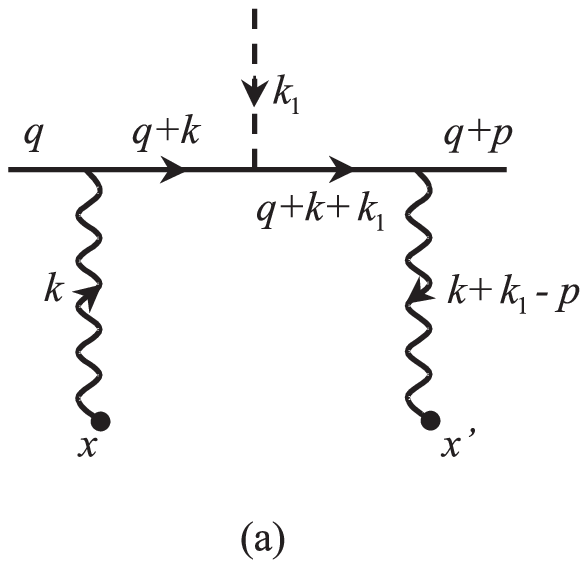}
\includegraphics{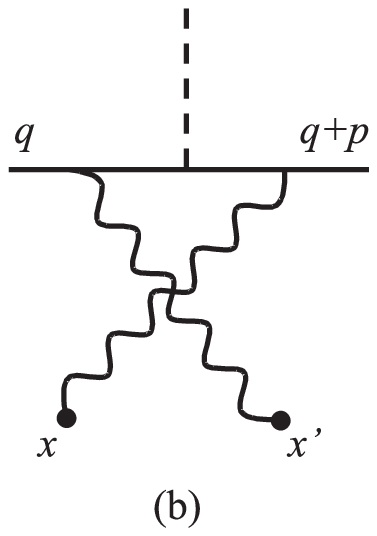}
\caption{Kazakov, Quantum fluctuations of Coulomb potential as a
source of flicker noise. The influence of heat bath} \label{fig3}
\end{figure}

Next, the particle propagator is also modified by the external
field. It is not difficult to see that for a sufficiently small
field strength, $\bm{E},$ this modification can be treated
perturbatively. It can be recalled that in coordinate space, it
amounts to multiplying the zero-field propagator $D^{\phi}(z,z')$
by\footnote{This phase factor incorporates contributions involving
only vertices linear in the photon field. This is sufficient for
the subsequent discussion concerned with the leading correction to
the propagator. Inclusion of the other interaction vertex adds
terms of higher orders in $\bm{E}$ to the exponent.}
$\exp\left\{ie(\bm{E},\bm{z}+\bm{z}')(z'_0-z_0)/2\right\}.$
Although the low-frequency limit is determined by the large-time
behavior of the quantities involved, implying that $(z'_0-z_0)\sim
1/\omega$ is large, the exponent can be made as small as desired
for any given $\omega$ by taking $|\bm{E}|$ sufficiently small.
This also will be clear from the explicit calculations to follow.
The lowest order correction to the correlation function is
represented by diagrams with a single insertion of the
particle-field interaction vertex into the internal solid line, as
shown in Fig.~\ref{fig3}. As in the preceding section, these
diagrams vanish unless all interaction vertices are 1-type, on
account of momentum conservation in the vertices together with the
mass shell conditions for the massive particle. The contribution
of the diagram \ref{fig3}(a) to the spectral density of the
two-point Green function
\begin{eqnarray}\label{diagenk1e}
G^E_a(\bm{x},\bm{x}',t',\omega) = && (4\pi e)^2\iiint \frac{d^3
\bm{q}}{(2\pi)^3} \frac{d^3 \bm{p}}{(2\pi)^3}\frac{d^3
\bm{k}_1}{(2\pi)^3}\frac{a(\bm{q})a^*(\bm{q} +
\bm{p})}{\sqrt{2\varepsilon_{\bm q}2\varepsilon_{{\bm q} +
\bm{p}}}} \nonumber\\&& \times e^{ip^0(t' - t_0) - i(\bm{p} -
\bm{k}_1)\bm{x}'}\varphi(\bm{k}_1)
J^{E}_a(p,q,\bm{x}-\bm{x}',\omega)\,,
\end{eqnarray}
\noindent where
\begin{eqnarray}\label{diagenk3e}&&
J^E_a(p,q,\bm{x}-\bm{x}',\omega) = - \left.ie\int \frac{d^3
\bm{k}}{(2\pi)^3} e^{i\bm{k}(\bm{x}'-\bm{x})}(2q_0 + k_0)(2q_0 +
2k_0)(2q_0 + k_0 + p_0)\right.\nonumber\\&& \left.
D_{11}(k)D^{\phi}_{11}(q+k)D^{\phi}_{11}(q+k+k_1)D_{11}(k+k_1-p)
\right|_{k_0=\omega}\,, \quad k_1 = (0,\bm{k}_1)\,,
\end{eqnarray}
\noindent and $\varphi(\bm{k})$ is the Fourier transform of the
external field potential, $\varphi(\bm{x}) = -(\bm{E},\bm{x}).$ We
do not include an arbitrary constant in this expression because it
is clear in advance that it cannot affect the final result. It is
proved in the Appendix that the correlation function is actually
invariant under the most general gauge variations of the
electromagnetic potential. Substituting
$$\varphi(\bm{k}) = - i(2\pi)^3 \left(\bm{E}\frac{\partial}{\partial
\bm{k}}\right)\delta^{(3)}(\bm{k})$$ into Eq.~(\ref{diagenk1e})
and integrating by parts, the $\bm{k}_1$ integral in
Eq.~(\ref{diagenk1e}) is brought to the form
\begin{eqnarray}\label{k1int}&&
i\int d^3\bm{k}_1 \delta^{(3)}(\bm{k})
\left(\bm{E}\frac{\partial}{\partial
\bm{k}_1}\right)\left[D^{\phi}_{11}(q+k+k_1)D_{11}(k+k_1-p)e^{-
i(\bm{p} - \bm{k}_1)\bm{x}'}\right]\,,
\end{eqnarray}
\noindent where only terms involving $\bm{k}_1$ are retained. As
we have seen, the singularity at $\omega = 0$ in the function
$G_a(\bm{x},\bm{x}',t',\omega)$ comes from integration over small
$\bm{k},$ and this singularity is now strengthened by the extra
particle propagator and the differentiation with respect to
$\bm{k}_1.$\footnote{Independently of the proof given in the
Appendix, it can be noticed that the constant term in the
potential does not involve the $\bm{k}_1$-differentiation, and
hence does not contribute to the leading singularity for $\omega
\to 0$ anyway.} Consider first the case when the
$\bm{k}_1$-derivative acts on the last two factors in the square
brackets. Since these depend on the difference
$(\bm{p}-\bm{k}_1),$ changing $\partial/\partial \bm{k}_1 \to -
\partial/\partial \bm{p},$ and then integrating by parts with respect
to $\bm{p}$ in Eq.~(\ref{diagenk1e}), this derivative is rendered
to act on terms independent of $\bm{k}$\footnote{Moreover, if the
function $a(\bm{q})$ is decomposed as in Eq.~(\ref{abrel}), and
the potential is chosen respectively to vanish at the point
$\bm{x}_0,$ i.e., $\varphi(\bm{x}) = -(\bm{E},\bm{x} - \bm{x}_0),$
then the $\bm{p}$-derivative acts on the product of
$e^{i\bm{p}\bm{x}_0}$ with the slowly varying factors
$b(\bm{q}+\bm{p}),$ $\varepsilon_{\bm{q}+\bm{p}},$ $p_0.$
Differentiation of the exponent gives $\bm{x}_0$ which just
cancels the purposely chosen constant term in the potential, while
the result of differentiation of the remaining factors can be
neglected to the leading order of the long-range expansion. This
observation can be useful in assessing the higher order
corrections to the correlation function.}. Thus, of the three
factors in the square brackets, only the first is to be
differentiated in effect, thus reducing the expression
(\ref{k1int}) to
$$\frac{2i(\bm{E},\bm{q} + \bm{k})}{[m^2 - (q+k)]^2}
D_{11}(k-p)e^{-i(\bm{p}\bm{x}')}\,.$$ Taking into account also the
vertex factor $e(2q_0 + 2k_0),$ we see that the first order
correction to the charged particle propagator due to constant
homogeneous external electric field is obtained by inserting the
factor
$$\frac{4ie(q_0+k_0)(\bm{E},\bm{q} + \bm{k})}{[m^2 - (q+k)]^2}
\equiv \varkappa $$ into the integrand in Eq.~(\ref{diagenk3}).
The rest of the calculation repeats the steps of
Sec.~\ref{zerofield}. Substituting explicit expressions for the
photon propagators, one sees that only the part proportional to
$\delta(k^2)$ is to be retained in the expression for $D_{11}(k),$
while the corresponding part in $D_{11}(k-p)$ is to be omitted.
One consequence of this observation is that the factor $\varkappa$
simplifies to $$\frac{ie(\bm{E}\bm{q})}{m\omega^2}\,,$$ where we
have taken into account that $k_0 = \omega \ll q_0,$ and $|\bm{k}|
= |\omega| \ll |\bm{q}|$ [indeed, taking our standard example of
electron in a crystal, the ratio $|\omega|/|\bm{q}|$ is, in the
ordinary units, $(\hbar|\omega|/c)/|\bm{q}| \sim |\omega|d/c\sim
|\omega|10^{-18},$ with $\omega$ expressed in Hz]. Another
consequence is that the contributions of diagrams \ref{fig3}(a),
\ref{fig3}(b) are related, as before, by\footnote{It is convenient
to prove this relation before the integration over $\bm{k}_1$
using the following sequence of substitutions: $\bm{k}\to \bm{k} +
\bm{p} - \bm{k}_1\,,$ $\bm{q} \to \bm{q} - \bm{p}\,,$ and then
$\bm{p} \to - \bm{p}\,.$ The extra factor $(-1)$ coming from
complex conjugation of the imaginary unit in the factor
$\varkappa$ is compensated by that from the integration by parts
with respect to $\bm{k}_1.$ Alternatively, this can be proved
directly in coordinate space (after extracting the relevant
contribution) by substituting
$\exp\left\{ie(\bm{E},\bm{z}+\bm{z}')(z'_0-z_0)/2\right\}D^{\phi}(z,z')$
for the particle propagator, interchanging the integration
variables $z,z',$ and taking into account symmetry of the
functions $D(x,z),$ $D^{\phi}(z,z')\,.$}
$$G^E_b(\bm{x},\bm{x}',t',\omega) =
G^{E*}_a(\bm{x},\bm{x}',t',-\omega)\,.$$ Furthermore,
Eq.~(\ref{diagenk4}) is now replaced by
\begin{eqnarray}\label{diagenk5e}
J^E_a(p,q,\bm{x}-\bm{x}',\omega) &=& \varkappa\frac{2\pi (2m +
p_0)}{\omega\beta |\omega| p^2}\int \frac{d^3 \bm{k}}{(2\pi)^3}
e^{i\bm{k}(\bm{x}'-\bm{x})}\delta(\omega^2 - \bm{k}^2 )
\nonumber\\
&=& \frac{ie(\bm{E}\bm{q})}{m\omega^2}\frac{(2m + p_0)}{2\pi
\omega\beta p^2}\left[1 - \frac{\omega^2}{6}(\bm{x}'-\bm{x})^2
\right] + O(\omega)\,.
\end{eqnarray}
\noindent An important difference in comparison with the result of
the preceding section is that because of the extra imaginary unit
brought in by the factor $\varkappa,$ the term proportional to
$p_0$ in the last formula gives rise now to a purely real
contribution upon substituting $J^E_a$ in Eq.~(\ref{diagenk1e}),
and hence is cancelled\footnote{Here $\bm{p}$ is neglected in
comparison with $\bm{q}.$ Otherwise, there is a residual term
proportional to $p_0$ in the total of the two diagrams. The usual
manipulation with the integration variables $\bm{q}\to \bm{q} -
\bm{p},$ $\bm{p}\to-\bm{p}$ shows that this term can be obtained
by substituting $(\bm{E}\bm{q})\to (\bm{E}\bm{p}).$ Its ratio to
the main contribution considered in the text is
$|\bm{p}|p_0/m|\bm{q}| \sim \bm{p^2}/m^2.$ For electrons in a
sample of characteristic size $l\,{\rm cm},$ this is, in the
ordinary units, $(\hbar/lmc)^2 \sim 10^{-20}/l^2.$} in the sum of
diagrams \ref{fig3}(a), \ref{fig3}(b). On the other hand, the
terms independent of $p_0$ survive, and lead to the following
expression for the first order correction to the spectral density
of the two-point Green function
\begin{eqnarray}\label{gefree}
G^E(\bm{x},\bm{x}',t',\omega) = - \frac{16\pi i
e^3}{m\beta\omega^3}\hspace{-0,4cm}&&\left[1 -
\frac{\omega^2}{6}(\bm{x}'-\bm{x})^2 \right]\nonumber\\&& \times
{\rm Re} \iint \frac{d^3 \bm{q}}{(2\pi)^3}\frac{d^3
\bm{p}}{(2\pi)^3} a(\bm{q})a^*(\bm{q}+\bm{p})
\frac{(\bm{E}\bm{q})}{\bm{p}^2} e^{ip_0(t'-t_0) -
i(\bm{p}\bm{x}')}\,.
\end{eqnarray}
\noindent In a many-particle system, this result is to be
expressed through the one-particle density matrix, which is
accomplished by replacing $a^*(\bm{q}')a(\bm{q}) \to
\varrho_0(\bm{q}',\bm{q}),$ where $\varrho_0$ is the momentum
space density matrix at the time instant $t_0.$ We recall that
$t_0$ denotes the instant at which the particle state $a(\bm{q})$
is prepared. It can be identified, for instance, as the moment the
charge carrier enters the sample, or escapes from a surface trap,
etc. The factor $e^{ip_0(t'-t_0)}$ in the integrand then realizes
time evolution of the density matrix from $t_0$ to $t'.$ Since
$p_0 = (\bm{q} + \bm{p})^2/2m - \bm{q}^2/2m,$ this is a free
evolution. The fact that the evolution of charge carriers in
solids is not actually free on macroscopic scales is not important
for the present consideration in which appearance of the
$1/\omega$-singularity is related to the effects of the medium on
the field propagators. In this respect, it is essentially
different from considerations of Ref.~\cite{kazakov2}, in which
time evolution of the particle wave packet was the central issue,
and, in particular, the requirement that the particle collisions
be elastic was important. To take into account particle collisions
in the present case, it is sufficient to consider them as
instantaneous. Then the interval $(t_0,t')$ is divided into a
sequence of short time intervals of order $\tau_f$ (the particle
mean free time), on each of which the density matrix evolves
freely as in Eq.~(\ref{gefree}), and changes abruptly at the
collision instants. Going through this sequence the density matrix
tends to the stationary statistical distribution function,
$\varrho(\bm{q}',\bm{q}),$ which is independent of the initial
particle state. What is important here is the sign of the
difference $(t'-t_0).$ Recall that $t'$ is a fixed time instant to
count off the time interval $\tau$ with respect to which the
correlation function is Fourier-transformed, and that each
particle has its own $t_0.$ This means that for a given $\omega,$
the system is observed during the time interval $(t' - \Delta t,
t' + \Delta t),$ where $\Delta t\sim 1/\omega,$ and $t_0$s are
distributed uniformly over this interval. The density matrix
evolves forward (backward) in time, if $t'>t_0$ ($t'<t_0$). But
time reversal involves inversion of particle momentum, and
therefore, the reciprocal contributions to the function
$G^E(\bm{x},\bm{x}',t',\omega)$ have opposite signs. To be more
specific, let $(t'-t_0)>0.$ Then the exponent $e^{ip_0(t'-t_0)}$
realizes a forward evolution of the density matrix, so that the
integral in Eq.~(\ref{gefree}) takes eventually the form
\begin{eqnarray}\label{contrf}
\iint \frac{d^3 \bm{q}}{(2\pi)^3}\frac{d^3 \bm{p}}{(2\pi)^3}
\varrho(\bm{q}+\bm{p},\bm{q}) \frac{(\bm{E}\bm{q})}{\bm{p}^2} e^{-
i(\bm{p}\bm{x}')}\,.
\end{eqnarray}
\noindent On the other hand, if $(t'-t_0)<0,$ then the density
matrix evolves backward. In momentum space, the initial state of
the reversed motion is represented by the amplitude
$\tilde{a}(\bm{q}) = a^*(-\bm{q}).$ Taking complex conjugate of
the integral in Eq.~(\ref{gefree}) (which does not change the
value of $G^E$ in view of the sign ``Re''), and changing the
integration variables $\bm{q}\to - \bm{q},$ $\bm{p}\to - \bm{p}$
gives in this case
$$ - \iint \frac{d^3 \bm{q}}{(2\pi)^3}\frac{d^3 \bm{p}}{(2\pi)^3}
\tilde{a}(\bm{q})\tilde{a}^*(\bm{q}+\bm{p})
\frac{(\bm{E}\bm{q})}{\bm{p}^2} e^{ip_0(t_0-t') -
i(\bm{p}\bm{x}')}\,.$$ After replacement
$\tilde{a}^*(\bm{q}+\bm{p})\tilde{a}(\bm{q}) \to
\tilde{\varrho}(\bm{q}+\bm{p},\bm{q}),$ where $\tilde{\varrho}$
plays the role of the momentum density matrix at the moment $t',$
the exponent $e^{ip_0(t_0-t')}$ governs forward evolution of this
state on the interval $(t',t_0),$ so that the above expression
takes the form
$$ - \iint \frac{d^3 \bm{q}}{(2\pi)^3}\frac{d^3 \bm{p}}{(2\pi)^3}
\varrho(\bm{q}+\bm{p},\bm{q}) \frac{(\bm{E}\bm{q})}{\bm{p}^2} e^{-
i(\bm{p}\bm{x}')}\,.$$ The density matrix here is the same as in
(\ref{contrf}), because the statistical distribution is
independent of the initial state. We see that reciprocal
contributions to the function $G^E(\bm{x},\bm{x}',t',\omega)$
cancel each other when summed over all particles in the system.
Thus, we arrive at the important conclusion that the total noise
intensity is independent of the number of particles, and remains
at the level of individual contribution. As was shown in
Ref.~\cite{kazakov2}, this conclusion is also true of the
disconnected part of the correlation function, though by virtue of
quite different reasons.

It is customary to further express the function
$\varrho(\bm{q}',\bm{q})$ via the real mixed distribution
function, $n(\bm{r},\bm{q}),$ according to
\begin{eqnarray}
\varrho(\bm{q}+\bm{p},\bm{q}) = \int d^3\bm{r}e^{i(\bm{p}\bm{r})}
n\left(\bm{r},\bm{q} + \frac{\bm{p}}{2}\right)\,. \nonumber
\end{eqnarray}
\noindent Probability distributions for the particle position in a
sample and for its momentum can be obtained by integrating
$n(\bm{r},\bm{p})$ over all $\bm{p}$ and the sample volume,
respectively. Using this in the expression (\ref{contrf}), and
substituting the latter into Eq.~(\ref{gefree}) yields
\begin{eqnarray}\label{diagenk6e}
G^E(\bm{x},\bm{x}',t',\omega) = - \frac{16\pi i
e^3}{m\beta\omega^3}\hspace{-0,4cm}&&\left[1 -
\frac{\omega^2}{6}(\bm{x}'-\bm{x})^2 \right]\nonumber\\&& \times
{\rm Re} \iiint \frac{d^3 \bm{q}}{(2\pi)^3}\frac{d^3
\bm{p}}{(2\pi)^3} d^3\bm{r} \frac{(\bm{E}\bm{q})}{\bm{p}^2}
n\left(\bm{r},\bm{q} + \frac{\bm{p}}{2}\right)e^{i\bm{p}(\bm{r} -
\bm{x}')}\,.
\end{eqnarray}
\noindent As we know, the first term in the square brackets in
this formula doest not contribute to the voltage power spectrum,
and can be omitted. Furthermore, after shifting $\bm{q}\to \bm{q}
- \bm{p}/2,$ and omitting the imaginary term proportional to
$(\bm{E},\bm{p}),$ the triple integral becomes purely real, so the
symbol ``${\rm Re}$'' can be omitted. Integrating then over
$\bm{p}$ with the help of the formula
$$\int\frac{d^3\bm{p}}{(2\pi)^3}e^{i(\bm{p}\bm{x})}\frac{4\pi}{\bm{p}^2}
= \frac{1}{|\bm{x}|}\,,$$ we thus obtain
\begin{eqnarray}\label{diagenk7e}
G^E(\bm{x},\bm{x}',t',\omega) = \frac{2i
e^3(\bm{x}'-\bm{x})^2}{3m\beta\omega} \iint \frac{d^3
\bm{q}}{(2\pi)^3}d^3\bm{r} (\bm{E}\bm{q})
\frac{n(\bm{r},\bm{q})}{|\bm{r}-\bm{x}'|}\,.
\end{eqnarray}
\noindent Equation~(\ref{corrg}) shows that the spectral density
of the correlation function is given by the same expression
(\ref{diagenk7e}), so that the power spectrum of voltage
fluctuations is, according to Eq.~(\ref{corrfup}),
\begin{eqnarray}\label{main}
C_U(\bm{x},\bm{x}',t',\omega) = - \frac{2i
e^3(\bm{x}'-\bm{x})^2}{3\beta\omega \Omega} \int d^3\bm{r}
(\bm{E}\overline{\bm{v}}(\bm{r}))\left( \frac{1}{|\bm{r}-\bm{x}|}
+ \frac{1}{|\bm{r}-\bm{x}'|}\right)\,,
\end{eqnarray}
\noindent where $$\overline{\bm{v}}(\bm{r}) = \Omega\int\frac{d^3
\bm{q}}{(2\pi)^3} \frac{\bm{q}}{m}n(\bm{r},\bm{q})$$ is the local
drift velocity of the charge carriers, $\Omega$ denoting the
sample volume. For a crystal in the homogeneous external field,
$\overline{\bm{v}}$ is a function of the crystalline direction,
$$\overline{v}_i = \mu_{ik}E_k\,, \quad i,k=1,2,3,$$ where $\mu_{ik}$
is the charge carrier mobility tensor. With the help of this
formula Eq.~(\ref{main}) can be rewritten as
\begin{eqnarray}\label{mainhom}
C_U(\bm{x},\bm{x}',t',\omega) = - i\eta\,\frac{U^2_0}{\omega}\,,
\quad \eta \equiv \frac{2e^3\mu g}{3\beta}\,,
\end{eqnarray}
\noindent where $$\mu=\mu_{ik}n_i n_k\,, \quad \bm{n} =
\frac{\bm{E}}{|\bm{E}|}\,,$$ $U_0 = |\bm{E}||\bm{x} - \bm{x}'|$ is
the bias applied to the sample (it is assumed that $\bm{E}
\parallel (\bm{x} - \bm{x}'),$ as is usually the case), and $g$ is
a geometrical factor
\begin{eqnarray}\label{gfactor}
g \equiv \frac{1}{\Omega}\int\limits_{\Omega} d^3\bm{r}\left(
\frac{1}{|\bm{r}-\bm{x}|} + \frac{1}{|\bm{r}-\bm{x}'|}\right)\,.
\end{eqnarray}
\noindent If Fourier transformation is defined in a purely real
form, i.e., as a decomposition in $\cos(\omega \tau),$
$\sin(\omega\tau),$ rather than in $e^{i\omega\tau},$ then the
spectral density is also real:
\begin{eqnarray}\label{mainhomr}
C_U(\bm{x},\bm{x}',t',\omega) = \eta\,\frac{U^2_0}{\omega}\,,
\quad \eta \equiv \frac{2e^3\mu g}{3\beta}\,,
\end{eqnarray}
\noindent

We mention for future reference that if the sample is an elongated
parallelepiped with the leads attached to its ends (as is usually
the case in practice), then the $g$-factor can be evaluated as
\begin{eqnarray}\label{gfactorapprox}
g \approx \frac{2}{l w h}\int\limits_{w}^{l} \frac{whd x}{x} =
\frac{2}{l}\ln \frac{l}{w}\,,
\end{eqnarray}
\noindent where $l,w,h$ denote the sample length, width and
thickness, respectively, and it is assumed that $h< w\ll l.$ We
note also that in the ordinary units, the dimensionless factor
$\eta$ reads
$$\eta = \frac{2e^3\mu g}{3\beta \hbar^2 c^3} =
\frac{2\alpha^2}{3ec}g\mu T\,,$$ $\alpha$ being the fine structure
constant. In particular, in the $CGS$ system of units,
\begin{eqnarray}\label{etaapprox}
\eta \approx 3.4\cdot 10^{-22} g\mu T\,,
\end{eqnarray}
\noindent where the absolute temperature $T$ is to be expressed in
$^{\circ}{\rm K}.$

\subsection{On the unboundedness of $1/f$-spectrum}\label{unbound}

In this section, a special feature of the derived expression for
the power spectrum, namely, its oddness in frequency, will be
discussed in connection with the problem of observed absence of
frequency limits of the $1/f$-law. As was mentioned in
Introduction, $1/f$ noise has been detected in a very wide
frequency band  $\sim 10^{-6}\,{\rm Hz}$ to $10^6\,{\rm Hz}.$ This
fact represents one of the essential difficulties for theoretical
explanation, because all physical mechanisms underlying existing
models of flicker noise work in much narrower subbands, and none
of the models suggested so far has been able to explain the
observed plenum of the $1/f$-spectrum.

On the other hand, existence of bounds on this spectrum is
generally believed to be necessary in order to guarantee
finiteness of the total noise power. There is a well-known
argument \cite{flinn} according to which these limits are actually
unnecessary when the flicker noise exponent $\gamma$ is strictly
equal to unity, because the logarithmic divergence of the total
power is not a problem in this case in view of the existence of
natural frequency cutoffs such as the inverse Planck time and
lifetime of Universe. However, this reasoning does not work for
$\gamma \ne 1,$ in which case divergence is a power of the cutoff.
At the same time, the results obtained above reconcile
unboundedness of $1/f$-spectrum with the requirements of
stationarity and finiteness of the total noise power in a quite
natural way. Indeed, using Eq.~(\ref{mainhom}) we find
$$\int\limits_{-\infty}^{+\infty}\frac{d\omega}{2\pi} C_U(\bm{x},\bm{x}',t',\omega)
e^{i\omega\tau} = \eta
U^2_0\int\limits_{-\infty}^{+\infty}\frac{d\omega}{2\pi}
\frac{\sin(\omega \tau)}{\omega} = \frac{\tau}{|\tau|}\frac{\eta
U^2_0}{2}\,.$$ More generally, if the spectrum $C_U\sim
1/f^{\gamma}$ is continued to negative $f$'s as an odd function,
then for any $0<\gamma<2$ the integral
$$\int\limits_{-\infty}^{+\infty}d\omega C_Ue^{i\omega\tau} \sim
\int\limits_{0}^{+\infty}df \frac{\sin(2\pi f \tau)}{f^{\gamma}}$$
is convergent in both limits $f \to 0$ and $f \to \infty.$ In
particular, the singular contribution to the voltage variance
(i.e. to the quantity $C_U(t,t')|_{\tau = 0}$) vanishes.

Since appearance of odd contributions to the power spectrum is
somewhat unusual in macroscopic fluctuation theory, let us discuss
it in more detail. Under stationary external conditions, the
voltage noise power spectrum (to be denoted below simply as
$C_U(t,t'),$ with the spatial arguments suppressed, for brevity)
must be independent of $t'.$ This is an expression of the noise
stationarity, or, using a term more suitable for the subsequent
discussion, time homogeneity with respect to the macroscopic
system. It is usually realized as the requirement that $C_U(t,t')$
be a function of the difference $t-t' \equiv \tau.$ Since
$C_U(t,t')$ is also symmetric with respect to the interchange $t
\leftrightarrow t',$ an immediate consequence of this is that it
is actually a function of $|\tau|,$ and hence the spectral density
is a real even function of frequency. It is important, on the
other hand, that time homogeneity is not necessarily exhibited by
individual contributions to the total voltage fluctuation,
whatever mechanism of flicker noise generation be. In particular,
this property evidently does not take place at the microscopic
level, i.e., with respect to elementary processes such as charge
carrier trapping, surface or grain boundary scattering, etc.
Stationarity of the macroscopic process emerges usually upon
summation over a large number of individual contributions, so that
this microscopic inhomogeneity turns out to be inconsequential.
However, this summation is not the only way to obtain a stationary
correlation function symmetric in $t,t'.$ Another possibility,
which is realized in the present paper, is that flicker noise may
be a one-particle phenomenon, in the sense that the entire effect
can be ascribed to elementary fluctuations produced by single
charge carriers. In this case the function $C_U(t,t')$ does not
have to depend solely on $|\tau|,$ and as the explicit
calculations of Sec.~\ref{calcul} show, it actually does not.  As
was mentioned above, elementary processes are inhomogeneous in
time, and hence the symmetry with respect to $t \leftrightarrow
t'$ imposes no restriction on the $\tau$-dependence of the
correlation function. The only remaining requirement, namely
reality of the correlation function, implies that contributions to
the spectral density must be real even, {\it or} imaginary odd
functions of frequency [Cf.~Eq.~(\ref{corrg})]. These two cases
correspond to the Fourier decomposition of the function
$C_U(t'+\tau,t')$ in $\cos(\omega\tau)$ and $\sin(\omega\tau),$
respectively, and describe the parts symmetric and antisymmetric
with respect to the difference of its time arguments. Finally,
transition to the statistical distribution removes the
$t'$-dependence of the power spectrum [Cf. transition from
Eq.~(\ref{gefree}) to Eq.~(\ref{diagenk6e})]. This restores
macroscopic time homogeneity of the correlation function, but
leaves the possibility of being odd with respect to the difference
of its time arguments. In other words, dependence of the power
spectrum on $t'$ shows itself only at microscopic scales, while
macroscopically fluctuations look as if they were homogeneous in
time.

The $1/f$-spectrum derived in the previous section has no lower
frequency cutoff. As to the upper bound, it is given by the
condition $f\ll T$ [see Sec.~\ref{prelim1}], or in the ordinary
units, $f\ll k T/\hbar \approx 10^{11}T\,{\rm Hz},$ with $T$
expressed in $^{\circ}{\rm K}.$ We see that from the practical
point of view, the obtained spectrum has no upper cutoff either.

\subsection{Validity of Eq.~(\ref{mainhomr}) and comparison with experimental data}
\label{comparison}

Let us next discuss the range of applicability of the obtained
results. The validity of the perturbative treatment of the
external field imposes a very strong bound on the field strength.
For this purpose we first collect all characteristic factors that
have appeared in the course of extracting the
$1/\omega$-asymptotic of the voltage power spectrum. As we have
seen, insertion of the vertex describing interaction of the
virtual charged particle with external field amounts to
multiplying the zero-field diagram by the factor, in the ordinary
units, $e(\bm{E}\bm{q})/(m\omega^2 \hbar)\,,$ which is eventually
promoted by the $\bm{q}$-integration into
$e\mu\bm{E}^2/(\omega^2\hbar).$ Furthermore, the leading
non-vanishing contribution to the voltage correlation function has
been obtained after expanding $e^{i\bm{k}(\bm{x} - \bm{x}')}$ in
the integrand of the $\bm{k}$-integral, which brought in a factor
$\omega^2(\bm{x} - \bm{x}')^2/c^2.$ Thus, the overall factor is
$e\mu\bm{E}^2(\bm{x} - \bm{x}')^2/(c^2\hbar) = e\mu
U^2_0/(c^2\hbar).$ This is small provided $\mu U^2_0\ll c^2\hbar/e
\approx 10^{3}$ units $CGS,$ which is a quite soft requirement met
in virtually all flicker noise measurements (Cf. examples below).
The problem with this estimation, however, is that at higher
orders, the scalar product $(\bm{E}\bm{q})$ is to be estimated as
$|\bm{E}||\bm{q}|,$ because $\overline{(\bm{E}\bm{q})^2}$ is of
the order $(|\bm{E}||\bm{q}|)^2\,,$ rather than
$(\bm{E}\overline{\bm{q}})^2\,.$ As a result, the requirement that
the factor $e(\bm{E}\bm{q})/(m\omega^2 \hbar)$ be small leads to
the following upper bound on the electric field strength for a
given frequency $\omega,$ in the $CGS$ system of units, $|\bm{E}|
\ll dm\omega^2/e\approx 10^{-25}\omega^2.$ At the same time, the
values $|\bm{E}|\sim 1$ are quite normal in flicker noise
measurements. In other words, from the point of view of the
developed theory, the experimentally relevant regime is identified
as the strong field limit. Yet the use of Eq.~(\ref{mainhomr}) in
this limit can be justified to a certain extent by recalling that
the perturbative expansion is in reality an asymptotic expansion,
and hence the fact that Eq.~(\ref{mainhomr}) gives the {\it first}
non-vanishing term of the voltage noise power spectrum implies
that the question of validity of the perturbative expansion is
actually a question of whether or not it is legitimate to use this
expansion to obtain higher order corrections to
Eq.~(\ref{mainhomr}). A rigorous justification is a difficult task
because it requires the use of non-perturbative methods. Thus,
this issue is left open until careful investigation of the strong
field limit. One of the possible ways this problem can hopefully
be resolved is a partial summation of the perturbation series,
followed by an analytical continuation with respect to
$\varkappa.$

After this discouraging observation of strong divergence of the
asymptotic series, the more striking turns out to be the fact that
Eq.~(\ref{mainhomr}) is in a general agreement, qualitative and
even quantitative, with the existing experimental data. First of
all, the spectral density is quadratic in the applied bias. This
is perhaps the most solidly established property of flicker noise.
Second, the noise level is inversely proportional to the sample
size. As to the dependence of flicker noise amplitude on sample
dimensions, agreement in the literature is not that good.
Experiments are usually arranged so as to prove one of the two
main competing points of view on the flicker noise origin, namely
wether it is a bulk or surface effect. Although this issue is far
from being resolved, there is no doubt that the noise level
increases with decreasing sample size. Third, it is generally
agreed that, with other things being equal, the flicker noise is
more intensive in semiconductors than in metals, and this is again
in conformity with Eq.~(\ref{mainhomr}), because charge carrier
mobility is higher in semiconductors than in metals, usually by
several orders. Unfortunately, determination of mobility in
semiconductors (or semimetals) is a difficult problem, both
theoretically and experimentally, and different experiments often
give significantly different results. By this reason, the
subsequent quantitative consideration will be carried out for
metals only. Even in this case careful estimation of the noise
level takes some effort. This is because electron mobilities in
thin metal films commonly used in flicker noise measurements
differ essentially from the corresponding bulk values, varying
non-monotonically with the film thickness, and exhibiting
complicated temperature dependence. Thus, the thicker the film,
the more reliable comparison of theoretical and experimental
results. Fortunately, the modern instrumentation allows
measurements in sufficiently thick samples, electrical transport
in which has bulk properties (usually, effects related to film
thickness become important for $h$ less than a few hundred
nanometers). As is well known, temperature dependence of the
electron mobility in this case is well approximated by the $1/T$
law. Theoretically, this approximation is valid for $T$ higher
than the Debay characteristic temperature, but in most cases it is
practically applicable already for $T\gtrsim 50^{\circ}{\rm K}.$
Thus, it follows from Eq.~(\ref{mainhomr}) that the flicker noise
level in thick samples is temperature independent. This conclusion
is confirmed, e.g., by the results of Ref.~\cite{massiha} where
$1/f$ noise was measured in $2.44\,{\rm \mu m}$ thick metal films,
which is quite sufficient for bulk treatment of the sample
conduction. According to Fig.~5 of Ref.~\cite{massiha}, the
flicker noise level is constant for $T\gtrsim 50^{\circ}{\rm K}$
indeed. Unfortunately, the authors of \cite{massiha} did not
specify the metals used in their experiments, which makes further
comparison with Eq.~(\ref{mainhom}) impossible.

In order to compare the absolute value of the noise spectral
density given by Eq.~(\ref{mainhomr}) with experimental data, we
use the results of the classic paper \cite{voss1} where flicker
noise in thin metal films was investigated. The information
provided in this paper is sufficient for estimation of the noise
intensity in the gold film shown in Fig.~2 of \cite{voss1}. This
was an elongated sample with $h = 25\,{\rm nm},$ $w = 8\,{\rm\mu
m},$ $l = 625\,{\rm\mu m},$ biased at $U_0 = 0.81\,{\rm V},$ and
operated at about $40^{\circ}{\rm K}$ above room temperature.
Substituting the sample dimensions in Eq.~(\ref{gfactorapprox})
gives $g = 140\,{\rm cm}^{-1}.$ Estimation of the electron
mobility is more subtle. As was mentioned above, charge carrier
mobility in thin films strongly deviates from its bulk value, and
this deviation is the main source of uncertainty in evaluating the
noise level. In the case under consideration, $\mu$ is isotropic
and can be found using the relation $\mu = \sigma/en,$ where
$\sigma$ is the electrical conductivity of gold, and $n = 5.9\cdot
10^{22}\,{\rm cm}^{-3}$ is the free electron concentration. The
bulk conductivity of gold at $T = 330^{\circ}{\rm K}$ is equal to
$4.0\cdot 10^{7}\,{\rm \Omega^{-1} m^{-1}},$ but in thin films the
value of $\sigma$ is strongly affected by the grain boundary and
surface scattering, surface roughness and other factors. The
relevant value of conductivity can be calculated indirectly using
the $I-V$ characteristic of the given gold sample, shown in Fig.~3
of \cite{voss1}. According to this figure, the sample resistance
was about $100\, \Omega.$ Taking into account the sample
dimensions given above, this implies that $\sigma =1.2\cdot
10^{6}\,{\rm \Omega^{-1} m^{-1}}.$ It should be mentioned that
this value is approximately six times lower than that obtained in
more recent studies of electrical transport in thin films. For
instance, according to Ref.~\cite{bieri} conductivity of a
$25\,{\rm nm}$ thick, $15\,{\rm \mu m}$ wide gold film obtained by
a laser-improved deposition of nanoparticle suspension, is $7.1
\cdot 10^{6}\,{\rm \Omega^{-1} m^{-1}}.$ The same value can be
obtained also indirectly using the data given in
Refs.~\cite{chen,pov}. According to \cite{chen}, the conductivity
of gold is $75\%$ to $85\%$ of its bulk value for $h = 100\,{\rm
nm},$ depending on the choice of the substrate, and decreases
below that value approximately linearly with decreasing thickness.
On the other hand, according to Ref.~\cite{pov} conductivity drops
to about $3\cdot 10^{5}\,{\rm \Omega^{-1} m^{-1}}$ for $h =
5\,{\rm nm}.$ One readily finds from this that for $h = 25\,{\rm
nm},$ $\sigma = (6.5 \div 7.5) \cdot 10^{6}\,{\rm \Omega^{-1}
m^{-1}}.$ Presumably, this difference in the values of
conductivity is to be attributed to the quality of film
deposition. In the case of $\sigma =1.2\cdot 10^{6}\,{\rm
\Omega^{-1} m^{-1}},$ the electron mobility equals to $\mu =
1.3\,{\rm cm^2/Vs},$ and then Eq.~(\ref{etaapprox}) gives $\eta =
6.0\cdot 10^{-15}\,.$ Substituting this together with the bias
value given above in Eq.~(\ref{mainhomr}), and setting $\omega =
2\pi f$ yields $C_U = 6.3\cdot 10^{-16}\,{\rm V^2/Hz}$ for the
frequency $f = 1\,{\rm Hz},$ which is to be compared with the
experimental value $C_U \approx 10^{-15}\,{\rm V^2/Hz}\,.$

\section{Discussion and conclusions}

We have shown that the combined action of the temperature and
external field effects results in appearance of a principally new
contribution to the power spectral density of quantum
electromagnetic fluctuations, given by Eqs.~(\ref{main}),
(\ref{mainhom}). The power spectrum is thus modified both
qualitatively and quantitatively. Being odd with respect to
frequency, the new term in the power spectrum describes
correlations in the values of voltage measured at two time
instants, which are finite for all times. The underlying reason
that makes the appearance of the new term possible (apart from the
two factors mentioned in the beginning of this paragraph) is the
inhomogeneity in time of fluctuations produced by individual
charge carriers. As discussed in Sec.~\ref{unbound}, oddness of
the found $1/f$-contribution gives a natural explanation to the
observed unboundedness of flicker noise spectrum. Although the
obtained result is valid, strictly speaking, only for very weak
fields, we have seen in Sec.~\ref{comparison} that it is in
qualitative and quantitative agreement with experimental data even
beyond its formal range of applicability.

Next, an important qualitative difference of the present
considerations from those of Ref.~\cite{kazakov1} is to be
emphasized. As we have seen in Sec.~\ref{calcul}, frequency
dependence of the power spectral function is determined completely
by internal structure of the Feynman diagrams representing the
connected part of the correlation function. In other words,
dispersion of the correlation function, considered in the present
paper, is related to the properties of virtual quanta propagation,
and not to the time evolution of the charge carrier wave function.
This is in contrast to considerations of Ref.~\cite{kazakov1}
where the $1/f$ asymptotic of the power spectrum was related to
the spreading of the particle wave packet, and was derived by
evaluating the disconnected part of the correlation function.

Finally, regarding discussion of Sec.~\ref{comparison} it should
be stressed that for the purpose of experimental verification of
Eq.~(\ref{mainhom}) only the genuine $1/f$ noise data was used,
i.e., the data that fits the law $f^{-\gamma}$ in which $\gamma =
1,$ within experimental error. Otherwise the comparison would be
meaningless, even for $f\approx 1\,{\rm Hz}.$ Large deviations of
$\gamma$ from unity, observed in some thin films, are presumably
due to back reaction of the conducting medium on the
electromagnetic field produced by the charge carriers. This issue
will be considered elsewhere.

\acknowledgments{I thank Drs. P.~I.~Pronin, G.~A.~Sardanashvili,
and K.~V.~Stepanyantz  (Moscow State University) for interesting
discussions.}

\begin{appendix}

\section{Gauge independence of the correlation function}

Consider the theory of interacting scalar and electromagnetic
fields described by the action
$$S = S_{\phi} + S_{A}\,,$$ where $S_{\phi}$ is given by
Eq.~(\ref{action}), and
$$S_{A} = - \frac{1}{4}{\displaystyle\int} d^4 x F_{\mu\nu} F^{\mu\nu}
+ S_{gf}\,, \quad S_{gf} = \frac{1}{2\alpha}{\displaystyle\int}
d^4 x~(\partial_{\mu}A^{\mu})^2\,,\quad F_{\mu\nu} =
\partial_{\mu} A_{\nu} - \partial_{\nu} A_{\mu}\,.$$ For arbitrary constant parameter
$\alpha,$ the gauge fixing term describes the generalized Lorentz
gauge. Let us introduce the generating functional of Green
functions
\begin{eqnarray}\label{gener1}
Z[J,\eta,\eta^*] = \int dA d\phi d\phi^* \exp\left\{i\left(S +
\int d^4x[J^{\mu}A_{\mu} + \eta^*\phi + \eta\phi^*]
\right)\right\}\,,
\end{eqnarray}
\noindent where $J,\eta,\eta^*$ denote sources for the fields
$A,\phi^*,\phi,$ respectively. Vanishing of $Z$ under the gauge
variation of the functional integral variables $$\delta A_{\mu} =
\partial_{\mu}\xi(x)\,, \quad \delta\phi = ie\xi(x)\phi\,,
\quad \delta\phi^* = -ie\xi(x)\phi^*,$$ with $\xi(x)$ a small
gauge function, leads to the Ward identity
\begin{eqnarray}\label{ward}
- i\partial_{\mu} J^{\mu}(y)Z +
\frac{\Box}{\alpha}\partial_{\mu}\frac{\delta Z}{\delta
J_{\mu}(y)} + ie\eta^*(y)\frac{\delta Z}{\delta \eta^*(y)} -
ie\eta(y)\frac{\delta Z}{\delta \eta(y)} = 0\,.
\end{eqnarray}
\noindent Since we are interested in the connected contribution to
the correlation function, we rewrite this identity for the
generating functional of connected Green functions, $W = - i\ln
Z,$
\begin{eqnarray}\label{wardc}
- \partial_{\mu} J^{\mu}(y) +
\frac{\Box}{\alpha}\partial_{\mu}\frac{\delta W}{\delta
J_{\mu}(y)} + ie\eta^*(y)\frac{\delta W}{\delta \eta^*(y)} -
ie\eta(y)\frac{\delta W}{\delta \eta(y)} = 0\,.
\end{eqnarray}
\noindent The consequence of this equation we need is obtained by
functional differentiation with respect to $\eta,\eta^*,$ and
twice with respect to $J,$ with all the sources set equal to zero
afterwards,
\begin{eqnarray}&&
\frac{\Box^y}{\alpha}\partial^y_{\mu}\frac{\delta^5 W}{\delta
J_{\mu}(y)\delta J_{\alpha}(x)\delta
J_{\beta}(x')\delta\eta(z)\delta\eta^*(z')} + ie\delta^{(4)}(y -
z')\frac{\delta^4 W}{\delta J_{\alpha}(x)\delta
J_{\beta}(x')\delta\eta(z)\delta \eta^*(y)} \nonumber\\&& -
ie\delta^{(4)}(y - z)\frac{\delta^4 W}{\delta J_{\alpha}(x)\delta
J_{\beta}(x')\delta \eta(y)\delta\eta^*(z')} = 0\,. \nonumber
\end{eqnarray}
\noindent Fourier transform of this identity with respect to $y$
reads
\begin{eqnarray}\label{wardc1}&&
\frac{k^2_1}{\alpha}k_{1\mu}\int d^4 y e^{-ik_1 y}\frac{\delta^5
W}{\delta J_{\mu}(y)\delta J_{\alpha}(x)\delta
J_{\beta}(x')\delta\eta(z)\delta\eta^*(z')} \nonumber\\ &&=
e(e^{-ik_1 z'} - e^{-ik_1 z}) \frac{\delta^4 W}{\delta
J_{\alpha}(x)\delta J_{\beta}(x')\delta\eta(z)\delta
\eta^*(z')}\,.
\end{eqnarray}
\noindent The argument of the Fourier transform is purposely
denoted here by $k_1$ to stress that the left hand side of this
equation corresponds to the variation of the Green function we
dealt with in Sec.~\ref{calcul}, under gauge variation of the
external field. Indeed, the longitudinal part of the photon
propagator in the generalized Lorentz gauge has the form
\begin{eqnarray}\label{long}
D^l_{\mu\nu}(k) = - \alpha\frac{k_{\mu}k_{\nu}}{k^4}\,.
\end{eqnarray}
\noindent Therefore, contraction with the factor $k^2_1
k_{1\mu}/\alpha$ is equivalent to amputation of the photon
propagator attached to the $y$ vertex, followed by contraction of
this vertex with $k_{1\mu}.$ Exactly the same result is obtained
under a gauge variation of the external field coming into this
vertex. The only difference with the Green function we considered
in Sec.~\ref{calcul} is that the external scalar lines in
Eq.~(\ref{wardc1}) are the particle propagators. To promote them
into particle amplitudes, according to the standard rules,
Eq.~(\ref{wardc1}) is to be Fourier transformed with respect to
the variables $z,z',$ and then multiplied by
$a(\bm{q})a^*(\bm{q}')(m^2 - q^2)(m^2 - q'^2),$ where the
arguments $q,q'$ of the Fourier transformations with respect to
$z,z'$ are to be taken eventually on the mass shell. But these
operations give zero identically when applied to the right hand
side of Eq.~(\ref{wardc1}), because each of the factors $e^{-ik_1
z'},$ $e^{-ik_1 z}$ makes the corresponding particle propagator
nonsingular on the mass shell. For instance, the first term in
Eq.~(\ref{wardc1}) gives rise to the contribution of the form
$(m^2 - q'^2)D^{\phi}(q' + k_1)$ times terms nonsingular on the
mass shell. For $k_1 \ne 0,$ the function $D^{\phi}(q' + k_1)$ is
also nonsingular at $q'^2=m^2,$ and hence this contribution
vanishes on the mass shell.

Thus, the correlation function is invariant under the gauge
transformations of the external field, which are part of the gauge
freedom in the theory. The other part is related to the explicit
dependence of the photon propagator on the choice of the gauge
conditions used to fix the gauge invariance of the action. As is
well known, it is the longitudinal part of the propagator that
depends on the gauge, and the most general Lorentz-invariant form
of this part is given by Eq.~(\ref{long}) in which $\alpha$ is to
be regarded as an arbitrary function of $k^2.$ It is not difficult
to see that variations of $\alpha(k^2)$ do not affect the
observable quantities. Recall, first of all, that we are
interested ultimately in the fluctuations of gauge-invariant
quantities such as the electric field strength. The
$\alpha$-independence of these quantities is a direct consequence
of their gauge invariance, because variations of $\alpha(k^2)$
give rise to terms that are pure gradients with respect to the
spacetime arguments $x,x',$ as is easily verified by substituting
the expression (\ref{long}) in place of one or two photon
propagators in Eq.~(\ref{diagen}). Then, if the vector potential
contribution to the field strength is negligible, as is the case
in our nonrelativistic calculation (recall the condition
$|\bm{q}|\ll m$ used throughout), the voltage correlation function
can be found by integrating the correlation function for the field
strength with respect to $\bm{x},\bm{x}'$ using the relation
$\bm{E} = - \bm{\nabla}A_0.$

Thus, gauge-independence of our results expressed by
Eqs.~(\ref{main}), (\ref{mainhom}) is proved.

\end{appendix}

\pagebreak

\centerline{\bf Figure captions}

Fig.1: Feynman diagrams representing connected part of correlation
function. Wavy lines denote photon propagators, solid lines
massive particle. $q$ and $p$ are the particle 4-momentum and
4-momentum transfer, respectively.

Fig.2: Symbolic diagrammatic picture of the effect of particle
collisions and external electric field (dashed line) on the
particle wave function.

Fig.3: Feynman diagrams describing the first order external field
correction to the particle propagator.

\pagebreak

\pagebreak

\pagebreak

\pagebreak

\end{document}